\documentclass[12pt]{article}
\begin{document}
      \sloppy

\def\AFOUR{%
\setlength{\textheight}{9.0in}%
\setlength{\textwidth}{5.75in}%
\setlength{\topmargin}{-0.375in}%
\hoffset=-.5in%
\renewcommand{\baselinestretch}{1.17}%
\setlength{\parskip}{6pt plus 2pt}%
}
\AFOUR
\def\car{\mathop{\square}}
\def\carre#1#2{\raise 2pt\hbox{$\scriptstyle #1$}\car_{#2}}

\parindent=0pt
\makeatletter
\def\section{\@startsection {section}{1}{\z@}{-3.5ex plus -1ex minus
   -.2ex}{2.3ex plus .2ex}{\large\bf}}
\def\subsection{\@startsection{subsection}{2}{\z@}{-3.25ex plus -1ex minus
   -.2ex}{1.5ex plus .2ex}{\normalsize\bf}}
\makeatother
\makeatletter
\@addtoreset{equation}{section}
\renewcommand{\theequation}{\thesection.\arabic{equation}}
\makeatother

\renewcommand{\a}{\alpha}
\renewcommand{\b}{\beta}
\newcommand{\g}{\gamma}           \newcommand{\G}{\Gamma}
\renewcommand{\d}{\delta}         \newcommand{\D}{\Delta}
\newcommand{\e}{\varepsilon}
\newcommand{\la}{\lambda}        \newcommand{\LA}{\Lambda}
\newcommand{\m}{\mu}
\newcommand{\A}{\widehat{A}^{\star a}_{\mu}}
\newcommand{\Ar}{\widehat{A}^{\star a}_{\rho}}
\newcommand{\n}{\nu}
\newcommand{\om}{\omega}         \newcommand{\OM}{\Omega}
\newcommand{\p}{\psi}             \newcommand{\PS}{\Psi}
\renewcommand{\r}{\rho}
\newcommand{\s}{\sigma}           \renewcommand{\S}{\Sigma}
\newcommand{\f}{{\phi}}           \newcommand{\F}{{\Phi}}
\newcommand{\vf}{{\varphi}}
\newcommand{\y}{{\upsilon}}       \newcommand{\Y}{{\Upsilon}}
\newcommand{\z}{\zeta}

\renewcommand{\AA}{{\cal A}}
\newcommand{\BB}{{\cal B}}
\newcommand{\CC}{{\cal C}}
\newcommand{\DD}{{\cal D}}
\newcommand{\EE}{{\cal E}}
\newcommand{\FF}{{\cal F}}
\newcommand{\GG}{{\cal G}}
\newcommand{\HH}{{\cal H}}
\newcommand{\II}{{\cal I}}
\newcommand{\JJ}{{\cal J}}
\newcommand{\KK}{{\cal K}}
\newcommand{\LL}{{\cal L}}
\newcommand{\MM}{{\cal M}}
\newcommand{\NN}{{\cal N}}
\newcommand{\OO}{{\cal O}}
\newcommand{\PP}{{\cal P}}
\newcommand{\QQ}{{\cal Q}}
\renewcommand{\SS}{{\cal S}}
\newcommand{\RR}{{\cal R}}
\newcommand{\TT}{{\cal T}}
\newcommand{\UU}{{\cal U}}
\newcommand{\VV}{{\cal V}}
\newcommand{\WW}{{\cal W}}
\newcommand{\XX}{{\cal X}}
\newcommand{\YY}{{\cal Y}}
\newcommand{\ZZ}{{\cal Z}}

\newcommand{\ch}{\widehat{C}}
\newcommand{\gh}{\widehat{\gamma}}
\newcommand{\W}{W_{i}}
\newcommand{\na}{\nabla}
\newcommand{\xint}{\dint d^4x\;}
\newcommand{\sla}{\raise.15ex\hbox{$/$}\kern -.57em}
\newcommand{\Sla}{\raise.15ex\hbox{$/$}\kern -.70em}
\def\h{\hbar}
\def\Lp{\displaystyle{\biggl(}}
\def\Rp{\displaystyle{\biggr)}}
\def\LP{\displaystyle{\Biggl(}}
\def\RP{\displaystyle{\Biggr)}}
\newcommand{\lp}{\left(}\newcommand{\rp}{\right)}
\newcommand{\lc}{\left[}\newcommand{\rc}{\right]}
\newcommand{\lac}{\left\{}\newcommand{\rac}{\right\}}
\newcommand{\identity}{\bf 1\hspace{-0.4em}1}
\newcommand{\complex}{{\kern .1em {\raise .47ex
\hbox {$\scriptscriptstyle |$}}
      \kern -.4em {\rm C}}}
\newcommand{\real}{{{\rm I} \kern -.19em {\rm R}}}
\newcommand{\rational}{{\kern .1em {\raise .47ex
\hbox{$\scripscriptstyle |$}}
      \kern -.35em {\rm Q}}}
\renewcommand{\natural}{{\vrule height 1.6ex width
.05em depth 0ex \kern -.35em {\rm N}}}
\newcommand{\tint}{\int d^4 \! x \, }
\newcommand{\intg}{\int d^D \! x \, }
\newcommand{\intm}{\int_\MM}
\newcommand{\tr}{{\rm {Tr} \,}}
\newcommand{\half}{\dfrac{1}{2}}
\newcommand{\pa}{\partial}
\newcommand{\pad}[2]{{\frac{\partial #1}{\partial #2}}}
\newcommand{\fud}[2]{{\frac{\delta #1}{\delta #2}}}
\newcommand{\dpad}[2]{{\displaystyle{\frac{\partial #1}{\partial
#2}}}}
\newcommand{\dfud}[2]{{\displaystyle{\frac{\delta #1}{\delta #2}}}}
\newcommand{\dfrac}[2]{{\displaystyle{\frac{#1}{#2}}}}
\newcommand{\dsum}[2]{\displaystyle{\sum_{#1}^{#2}}}
\newcommand{\dint}{\displaystyle{\int}}
\newcommand{\eg}{{\em e.g.,\ }}
\newcommand{\Eg}{{\em E.g.,\ }}
\newcommand{\ie}{{{\em i.e.},\ }}
\newcommand{\Ie}{{\em I.e.,\ }}
\newcommand{\nb}{\noindent{\bf N.B.}\ }
\newcommand{\etal}{{\em et al.}}
\newcommand{\etc}{{\em etc.\ }}
\newcommand{\via}{{\em via\ }}
\newcommand{\cf}{{\em cf.\ }}
\newcommand{\twiddle}{\lower.9ex\rlap{$\kern -.1em\scriptstyle\sim$}}
\newcommand{\qed}{\vrule height 1.2ex width 0.5em}
\newcommand{\grad}{\nabla}
\newcommand{\bra}[1]{\left\langle {#1}\right|}
\newcommand{\ket}[1]{\left| {#1}\right\rangle}
\newcommand{\vev}[1]{\left\langle {#1}\right\rangle}

\newcommand{\equ}[1]{(\ref{#1})}
\newcommand{\eq}{\begin{equation}}
\newcommand{\eqn}[1]{\label{#1}\end{equation}}
\newcommand{\eea}{\end{eqnarray}}
\newcommand{\eqa}{\begin{eqnarray}}
\newcommand{\eqan}[1]{\label{#1}\end{eqnarray}}
\newcommand{\ba}{\begin{array}}
\newcommand{\ea}{\end{array}}
\newcommand{\eqac}{\begin{equation}\begin{array}{rcl}}
\newcommand{\eqacn}[1]{\end{array}\label{#1}\end{equation}}
\newcommand{\qq}{&\qquad &}
\renewcommand{\=}{&=&} 
\newcommand{\cb}{{\bar c}}
\newcommand{\mn}{{\m\n}}
\newcommand{\pic}{$\spadesuit\spadesuit$}
\newcommand{\?}{{\bf ???}}
\newcommand{\Tr }{\mbox{Tr}\ }
\newcommand{\adot}{{\dot\alpha}}
\newcommand{\bdot}{{\dot\beta}}
\newcommand{\gdot}{{\dot\gamma}}

\global\parskip=4pt
\titlepage  \noindent
{
   \noindent

\hfill GEF-TH-20/2007 


\vspace{2cm}

\noindent
{\bf
{\large On Consistency Of Noncommutative Chern--Simons Theory
}}

\vspace{.5cm}
\hrule

\vspace{2cm}

\noindent
{\bf 
Alberto Blasi and Nicola Maggiore}
\footnote{alberto.blasi@ge.infn.ge, nicola.maggiore@ge.infn.it,}

\noindent
{\footnotesize {\it
 Dipartimento di Fisica -- Universit\`a di Genova --
via Dodecaneso 33 -- I-16146 Genova -- Italy and INFN, Sezione di
Genova 
} }

\vspace{1cm}
\noindent
{\tt Abstract~:}
We consider the noncommutative extension of Chern-Simons theory. We 
show the the theory can be fully expanded in power series of the 
noncommutative parameter $\theta$ and that no non-analytical sector exists. 
The theory appears to be unstable under radiative corrections, but we 
show that the infinite set of instabilities, to all orders in $\hbar$ 
and in $\theta$, is confined to a BRS exact cocycle. We show also 
that the theory is anomaly free. The 
quantum theory cannot be written in terms of the Groenewald-Moyal 
star product, and hence doubts arise on the interpretation of the noncommutative
nature of the underlying spacetime. Nonetheless, the deformed 
theory is well defined as a quantum field theory, and 
the beta function of the Chern-Simons coupling constant vanishes, as in the 
ordinary Chern-Simons theory. 
\vfill\noindent
{\footnotesize {\tt Keywords:}
BRST Quantization,
Chern-Simons Theories,
Topological Field Theories,
Noncommutative Field Theories.
\\
{\tt PACS Nos:} 
03.70.+k Theory of Quantized Fields,
11.15.-q Gauge Field Theories,
11.10.Gh Renormalization,
11.10.Nx Noncommutative Field Theory.
}
\newpage
\begin{small}
\end{small}

\setcounter{footnote}{0}


\section{Introduction}

It is common wisdom that a noncommutative extension of a quantum field 
theory can be realized by replacing the standard commutative product with 
the Groenewald-Moyal star product \cite{Douglas:2001ba,Szabo:2001kg}. 
When applied to gauge theories, 
this approach should worry about the fate of all the symmetries which, 
in the commutative case, are taken as the very definition of the model 
and which allow a complete discussion of its renormalizability  by 
analyzing the integrated cohomology of the BRS operator in the ghost 
charge zero sector (stability) and ghost charge one sector (anomaly).
We would like to assume this point of view from the very beginning and 
consider a noncommutative quantum gauge field model to be defined by 
its symmetries, locality and power counting. Of course the presence of 
the $\theta_{\mu\nu}$ parameter with inverse square mass dimension deeply 
affects the results of the cohomology analysis; in a previous 
investigation \cite{Blasi:2005bk,Blasi:2005vf},
where we have applied this method to the two 
dimensional BF model whose action is expanded in power series of 
$\theta$, we found it to be unstable already at first order in 
$\theta$ and also 
showed that the Groenewald-Moyal extension is not the general solution 
of the symmetry constraints.
In this paper we discuss in the same framework  the noncommutative 
Chern-Simons model, whose action has a  necessarily analytic expansion 
in the $\theta$ parameter.
We are able to carry out the analysis to all orders and the Chern-Simons 
model turns out to be much more robust than the BF one with respect to 
noncommutative deformations.
Indeed, the model is unstable in the sense that to any 
fixed order in 
$\theta$, the classical action acquires  new contributions with new free 
parameters, but these contributions never belong to the BRS cohomology, 
$i.e.$ are BRS variations. Even more important, the theory stays anomaly 
free to all orders.
Now, the absence of anomaly implies that the symmetry is maintained in 
the full noncommutative extension and the fact that the new 
contributions to the action are BRS trivial leads to the consideration 
that the noncommutative model can still be regarded as 
``renormalizable''
in a wider sense \cite{Gomis:1995jp,Blasi:1998ph},
since all new parameters belong to the nonphysical 
sector of the theory. Needless to say, the Groenewald-Moyal star product 
does not coincide with the general extension we propose here.
The plan of the paper is as follows: 
in Section 2  we briefly recall the defining symmetries of the classical model, 
the noncommutative Groenewald-Moyal extension and set up the tools to analyze 
the quantum theory to all orders in $\theta$.
Section 3 is devoted to the explicit discussion of the stability and anomaly 
problem up to the second order in $\theta$; the computation, whose feasibility 
relies on a previous result on the general solution of the linear vector 
supersymmetry \cite{Blasi:2006gq}, gives us a hint of what might be the all 
order result, which is proven in detail in the Appendix.  
Our conclusive considerations are collected in Section 4.

\section{The Classical Model}

The ordinary, commutative, Chern-Simons theory reads
\eq
S_{CS}=\frac{k}{2}\ \tr \int d^{3}x\ \epsilon^{\mu\nu\rho} \left (
{\bf A_{\mu}}\partial_{\nu}{\bf A_{\rho}}
-i\frac{2}{3} 
{\bf A_{\mu} A_{\nu} A_{\rho}}
\right )\ ,
\eqn{2.1}
where ${\bf A_{\mu}}\equiv T^{a}A^{a}_{\mu}$ and
the trace must be done on the group generators, which, for $SU(n)$, obey
\begin{eqnarray}
    \Tr(T^{a}T^{b}) &=& \delta^{ab} \label{2.2}\\
    \left[ T^{a},T^{b}\right ] &=& if^{abc}T^{c} \label{2.3} \\
    \{T^{a},T^{b}\} &=& d^{abc}T^{c}+\frac{1}{n}\d^{ab} \label{2.4}
    \end{eqnarray}
The action \equ{2.1} is invariant under the nilpotent BRS transformations
\begin{eqnarray}
    s A^{a}_{\mu} &=& -(D_{\mu}c)^{a}
    \equiv -(\partial_{\mu}c^{a} + f^{abc}A^{b}_{\mu}c^{c}) \nonumber \\
    s c^{a} &=& +\frac{1}{2}f^{abc}c^{b}c^{c} \label{2.5}\\
    s \bar{c}^{a} &=& b^{a} \nonumber\\
    s b^{a} &=&  0\ ,\nonumber
\end{eqnarray}    
where the fields $c^{a}(x)$, $\bar{c}^{a}(x)$ and $b^{a}(x)$
represent ghost, antighost and Lagrange multiplier, 
respectively, and belong to the adjoint 
representation of the  gauge group.    

The gauge fixing term is
\begin{eqnarray}
S_{gf} &=& s \int d^{3}x\ \bar{c}^{a}\partial^{\mu}A^{a}_{\mu} 
\nonumber \\
&=&
\int d^{3}x\ \left (
b^{a}\partial A^{a} +\bar{c}^{a}\partial^{\mu}(D_{\mu}c)^{a}
\right)\ . \label{2.6}
\end{eqnarray}
Notice that, in three dimensions the gauge parameter being massive, 
the Landau gauge choice is mandatory.

Once gauge fixed, the action
\eq
S=S_{CS}+S_{gf}
\eqn{2.7}

is invariant under an additional vector symmetry \cite{Delduc:1989ft}
\eq
\d_{\m}S=0\ ,
\eqn{2.8}
where
\begin{eqnarray}
    \delta_{\mu} A^{a}_{\nu} &=& 
    \frac{1}{k}\epsilon_{\mu\nu\rho}\partial^{\rho}\bar{c}^{a}  \nonumber \\
    \delta_{\mu} c^{a} &=& -A^{a}_{\mu} \label{2.9}\\
    \delta_{\mu} \bar{c}^{a} &=& 0 \nonumber\\
    \delta_{\mu} b^{a} &=& \partial_{\mu}\bar{c}^{a}  \ .\nonumber
\end{eqnarray}
The vector symmetry \equ{2.9} is peculiar to all topological field 
theories, and the algebra formed with the BRS operator is
\begin{eqnarray}
    s^{2} &=& 0 \label{2.10} \\
    \{\d_\m,\d_{\n}\} &=& 0 \label{2.11} \\
    \{ s,\d_{\m}\} &=& \partial_{\m} + \mbox{eqs of motion.} 
    \label{2.12}
    \end{eqnarray}
This algebraic structure, like the ordinary global supersymmetry, 
closes on translations, and plays a crucial role in the proof of 
finiteness of Chern-Simons theory, and of topological quantum field 
theories in general~\cite{Maggiore:1991aa,Birmingham:1991rh}.

Besides BRS symmetry \equ{2.5} and supersymmetry \equ{2.9}, the 
action \equ{2.7} 
shares with all gauge field theories built in the Landau gauge, the ghost 
equation \cite{Blasi:1990xz}
\eq
   \int d^{3}x\ \left(
    \fud{}{c^{a}} + f^{abc}\bar{c}^{b}\fud{}{b^{c}} \right ) S
    \equiv
     {\cal G}^{a}S=0\ .
    \eqn{2.13}

    In order to proceed towards the noncommutative extension of 
    Chern-Simons theory, it is customary to deform 
    the ordinary product between 
    quantum fields, into the Groenewald - Moyal 
    ``star'' product \cite{Douglas:2001ba,Szabo:2001kg}
\eq
\phi(x)\psi(x) \longrightarrow \phi(x)*\psi(x) \equiv
\lim_{y\rightarrow x}\
\exp(\frac{i}{2}\theta^{\mu\nu}\partial^{x}_{\mu}\partial^{y}_{\nu})\
\phi(x)\psi(y)\ ,
\eqn{2.14}
where $\theta^{\mu\nu}$ is a rank-two antisymmetric matrix which 
controls the noncommutative nature of spacetime coordinates
\eq
[x^{\mu},x^{\nu}]=i\theta^{\mu\nu}\ .
\eqn{2.15}    
Consequently, the noncommutative Chern-Simons (NCCS) action reads
\eq
S_{NCCS} = \frac{k}{2}\ \Tr \int d^{3}x\ \epsilon^{\mu\nu\rho} \left (
{\bf A}_{\mu}*\partial_{\nu}{\bf A}_{\rho}
- i\frac{2}{3}
{\bf A}_{\mu} * {\bf A}_{\nu} * {\bf A}_{\rho}
\right ) \ .
\eqn{2.16}
The action \equ{2.16} can be fully expanded in power series of 
$\theta$
\eq
S_{NCCS} = \sum_{n=0}^{\infty}S_{NCCS}^{(n)}\ ,
\eqn{2.17}
where
\begin{eqnarray}
    S_{NCCS}^{(0)} &=& S \label{2.18} \\
    S_{NCCS}^{(1)} &=& \theta^{\a\b}S_{\a\b}   \label{2.19} \\    
    S_{NCCS}^{(2)} &=&  \theta^{\a\b}\theta^{\g\d}S_{\a\b\g\d}\ ,  
    \label{2.20}  
\end{eqnarray}
and so on at higher orders. Up to second order in $\theta$,
$S$ is given by Eq. \equ{2.7}, and
\begin{eqnarray}
S_{\a\b} &=& \frac{1}{2}d^{abc}
\int d^{3}x\
\partial_{\a}A^{a}_{\m}\left (
\frac{1}{6}\e^{\m\n\r}
 \partial_{\b}A^{b}_{\n} A^{c}_{\r}
-
\partial^{\m}\bar{c}^{b}
\partial_{\b}c^{c}
\right) \label{2.21} \\
S_{\a\b\g\d} &=& -\frac{1}{8}f^{abc}
\int d^{3}x\
\partial_{\a\g}A^{a}_{\m}
\left(
\frac{1}{6} \e^{\m\n\r}
\partial_{\b\d}A^{b}_{\n}
A^{c}_{\r}
+
\partial^{\m}\bar{c}^{b}
\partial_{\b\d}c^{c}
\right) \label{2.22}
\end{eqnarray}
Accordingly, always up to $O(\theta^{2})$, the noncommutative BRS symmetry is
\begin{eqnarray}
    s^{(\theta)} A^{a}_{\mu} &=& s A^{a}_{\mu}
    -\frac{1}{2}\theta^{\a\b}d^{abc}\partial_{\a}A^{b}_{\m}\partial_{\b}c^{c}
    +\frac{1}{8}\theta^{\a\b}\theta^{\g\d}f^{abc}\partial_{\a\g}A^{b}_{\m}\partial_{\b\d}c^{c}
    \nonumber \\
    s^{(\theta)} c^{a} &=& s c^{a}
    +\frac{1}{4}\theta^{\a\b}d^{abc}\partial_{\a}c^{b}\partial_{\b}c^{c}
    -\frac{1}{16}\theta^{\a\b}\theta^{\g\d}f^{abc}\partial_{\a\g}c^{b}
    \partial_{\b\d}c^{c}\label{2.23}\\
    s^{(\theta)} \bar{c}^{a} &=&  s \bar{c}^{a} \nonumber\\
    s^{(\theta)} b^{a} &=&   s b^{a}\ .\nonumber
\end{eqnarray}
On the other hand, the supersymmetry $\d_{\m}$, being linear in the 
quantum fields, is not affected by the noncommutative extension
\eq
\d^{(\theta)}_{\m}=\d_{\m}\ .
\eqn{2.24}
As a nontrivial property, it can be verified that the noncommutative deformation shares with 
the ordinary theory the symmetries
\begin{eqnarray}
    s^{(\theta)}S_{NCCS} &=& 0 \label{2.25} \\
    \d_{\m} S_{NCCS}  &=& 0 \label{2.26} \\
    {\cal G}^{a}S_{NCCS} &=& 0 \label{2.27}
    \end{eqnarray}
and the algebraic structure
\begin{eqnarray}
    \left ( s^{(\theta)}\right )^{2} &=& 0 \label{2.28} \\
    \{\d_{\m},\d_{\n}\} &=& 0 \label{2.29} \\
    \{ s^{(\theta)},\d_{\m}\} &=& \partial_{\m} + \mbox{eqs of 
    motion}\ . \label{2.30}
    \end{eqnarray}
    
Two remarks are in order. 

The first concerns the choice of the gauge group of 
noncommutative gauge field theories, which is known that it should be
$U(n)$~\cite{Douglas:2001ba,Szabo:2001kg}. 
The reason, is that the gauge group $U(n)$ is closed under 
the star product while $SU(n)$, for instance, is not. This 
restriction does not reveal itself in the noncommutative action, because of 
the trace which is done on the group generators, which are traceless. 
But it is evident when composite operators are considered, like 
 $s^{(\theta)}{\bf A_{\m}}$ and $s^{(\theta)}{\bf c}$ in the 
 BRS transformations. 
These expressions, 
 indeed, involve the anticommutator \equ{2.4}, which does not form an 
 algebra, due to the central term $\frac{1}{n}\d^{ab}$,
 and are meaningful only for $U(n)$ gauge groups, for which the 
 central term disappears
 \eq
  U(n)\ :\ \{T^{a},T^{b}\} = d^{abc}T^{c}\ .
  \eqn{2.31}
More subtle is the necessity of $U(n)$ gauge groups to verify, for 
instance, the nilpotency of the noncommutative BRS operator 
$s^{(\theta)}$ \equ{2.28}, order by order in $\theta$. Nilpotency is 
achieved only thanks to the following nontrivial relation between structure 
constants and completely symmetric tensors
\eq
f^{abp}f^{cdp} =
d^{acp}d^{bdp} - d^{bcp}d^{adp}
\eqn{2.32}
which holds  for $U(n)$ groups only \cite{{MacFarlane:1968vc}}.

The second remark concerns the analyticity of the theory. The quantum 
action $\G_{NCCS}$ does not have any non-analytic sector in $\theta$. 
Indeed, let 
us suppose that the NCCS quantum action $\G_{NCCS}$ contains a sector 
which can be expanded in negative powers of $\theta$
\eq
\left. \G_{NCCS}\right |_{non\ analytic}
=\sum_{n=1}^{\infty}\frac{1}{\theta^{n}}\G_{NCCS}^{(n)}\ .
\eqn{2.33}
recalling that $\frac{1}{\theta}$ has mass dimensions $+2$, power 
counting implies that the mass dimension of $\G_{NCCS}^{(n)}$ is
\eq
\mbox{dim}\ (\G_{NCCS}^{(n)}) = 3-2n\ ,
\eqn{2.34}
which, of course, must be a non-negative quantity.
Thus, at most there is only one possible term in the non-analytic 
expansion \equ{2.33}
\eq
\left. \G_{NCCS}\right |_{non\ analytic}
=\frac{1}{\theta}\G_{NCCS}^{(1)}\ ,
\eqn{2.35}
but no such $\G_{NCCS}^{(1)}$, with mass dimension $+1$, 
can be constructed which is a color 
singlet and gauge invariant. 

Hence, the $\theta$-expansion of NCCS theory does not admit a 
non-analytical sector.

\section{The Quantum Extension}

Once we got rid of the non-analytical sector, the counterterm 
$\S_{c}^{(\theta)}$ can be fully expanded in power series of $\theta$:
\eq
\S_{c}^{(\theta)} = \sum_{n=0}^{\infty} \theta^{n}\cdot \S_{c}^{(n)}
\eqn{3.1}
In \equ{3.1}, the ``dot'' product denotes all possible ways to contract 
Lorentz indices in order to form a scalar quantity. For example
\begin{eqnarray}
    \theta \cdot \S_{c}^{(1)} &\equiv& 
    \theta_{\m\n}\S^{(1)\m\n} \label{3.2}\\
     \theta^{2} \cdot \S_{c}^{(2)} &\equiv& 
    \theta_{\m\n}\theta^{\m\n}\S^{(2)} +
    \theta_{\m\lambda}\theta^{\lambda}_{\n} \S^{(2)\m\n} +
    \theta_{\m\n}\theta_{\r\s}\S^{\m\n\r\s}\label{3.3} 
    \end{eqnarray}
    
In order that the action is stable under radiative corrections, the 
counterterm must obey the following constraints\footnote{We omit to 
introduce external fields to define the nonlinear BRS variations in 
\equ{2.23}. It is readily seen indeed, that their presence, 
not altering at all our results, would make the treatment much heavier. 
Therefore,
without loss of generality, also at the quantum level we 
shall continue to deal with the BRS operator, and not with the 
Slavnov-Taylor identity~\cite{Piguet:1995er}.}
\begin{eqnarray}
    {\cal G}^{a}\S_{c}^{(\theta)} &=& 0 \label{3.4}\\
    s^{(\theta)}\S_{c}^{(\theta)} &=& 0 \label{3.5}\\
    \d_{\m}\S_{c}^{(\theta)} &=& 0\label{3.6}
    \end{eqnarray}
Recalling that the noncommutative BRS operator can be 
$\theta$-expanded
\eq
s^{(\theta)}=s^{(0)} + s^{(1)} + s^{(2)}+\ldots
\eqn{3.7}
order by order in $\theta$, the stability equation \equ{3.5} reads, up 
to $O(\theta^{2})$
\begin{eqnarray}
    O(\theta^{0})  &:& s^{(0)}\S_{c}^{(0)} = 0 \label{3.8}\\
     O(\theta^{1})  &:& s^{(0)}\S_{c}^{(1)} +
     s^{(1)}\S_{c}^{(0)}= 0 \label{3.9}\\
 O(\theta^{2})  &:& s^{(0)}\S_{c}^{(2)} +
     s^{(1)}\S_{c}^{(1)} +  s^{(2)}\S_{c}^{(0)} = 0\label{3.10} 
     \end{eqnarray}
while the ghost equation \equ{3.4} and the supersymmetry constraint 
\equ{3.6}, which do not mix the $\theta$-sectors, hold at each order
\begin{eqnarray}
    {\cal G}^{a}\S_{c}^{(n)} &=& 0 \label{3.11}\\
    \d_{\m}\S_{c}^{(n)} &=& 0\ .\label{3.12}
    \end{eqnarray}
For the analysis of the quantum extension of the theory, it is 
extremely helpful to know the general solution of the supersymmetry equation
\eq
\d_{\m}X^{p}_{q} =0\ ,
\eqn{3.13}
where $p$ and $q$ denote respectively 
mass dimension and ghost number of the functional $X$. In 
\cite{Blasi:2006gq}, it 
has been proven that the most general solution of \equ{3.13} is
\eq
X^{p}_{q} = \e^{\m\n\r}\d_{\m}\d_{\n}\d_{\r}X^{p-3}_{q+3} 
\equiv \d^{3} X^{p-3}_{q+3}\ .
\eqn{3.14}
Now, $\S_{c}^{(0)}$ is a local integrated functional with mass 
dimensions $+3$ and ghost number $0$. According to Eq. \equ{3.12} and Eq. 
\equ{3.14}, it must be written
\eq
\S_{c}^{(0)} = \d^{3} X^{0}_{3}\ .
\eqn{3.15}
Since the ghost operator ${\cal G}^{a}$ \equ{2.13} anticommutes with both 
the BRS and the supersymmetry operators
\eq
\{{\cal G}^{a},s^{(\theta)}\}\ =\ 
\{{\cal G}^{a},\d_{\m}\}\ =0\ ,
\eqn{3.16}
it must also be
\eq
{\cal G}^{a}X^{0}_{3} =0\ ,
\eqn{3.17}
but no functional with the correct quantum numbers exists, hence
\eq
X^{0}_{3} =0\ ,
\eqn{3.18}
and
\eq
\S_{c}^{(0)}= 0\ .
\eqn{3.19}
We recovered here in a few lines a result which is already known, 
concerning the finiteness of commutative Chern-Simons theory 
\cite{Blasi:1989mw,Delduc:1990je}. 
This same technique easily leads us to get new results at 
higher orders in $\theta$.

Taking into account \equ{3.19}, at the first order in $\theta$, 
$\S_{c}^{(1)}$ must obey
\begin{eqnarray}
    s^{(0)}\S_{c}^{(1)} &=& 0 \label{3.20}\\
    {\cal G}^{a}\S_{c}^{(1)} &=& 0 \label{3.21}\\
    \d_{\a} \S_{c}^{(1)} &=& 0\label{3.22}
    \end{eqnarray}
From Eq. \equ{3.22}, we have
\eq
\S_{c}^{(1)} = \d^{3} \S^{dim=2}_{\Phi\Pi=3}\ ,
\eqn{3.23}
since, in order that $\theta\cdot\S_{c}^{(1)}$ has mass dimension 
$+3$, $\S_{c}^{(1)}$ must have dimension $+5$. In general
\eq
\mbox{dim}(\S_{c}^{(n)}) = 3 + 2n\ .
\eqn{3.24}
The only possible term, satisfying also the ghost condition \equ{3.21}, is
\eq
\S^{2}_{3\m\n} = \int d^{3}x \
d^{abc}c^{a}\partial_{\m}c^{b}\partial_{\n}c^{c}\ ,
\eqn{3.25}
which is not BRS invariant
\eq
s^{(0)}\S^{2}_{3\m\n}\neq 0\ ,
\eqn{3.26}
and therefore also
\eq
s^{(0)}\d\S^{(1)}_{\m\n}\neq 0\ ,
\eqn{3.27}
since, on integrated functionals,
\eq
\{s^{(0)},\d\}=0\ .
\eqn{3.28}
Hence, $\S^{(1)}_{\m\n}$ is ruled out by the symmetry constraints, and
\eq
\S^{(1)}_{c}=0\ .
\eqn{3.29}
Therefore, the NCCS theory, at least at first order in $\theta$, not only is 
stable under radiative corrections, but, more than that, keeps the 
property of finiteness displayed by the commutative theory.

At the next order in $\theta$, taking into account the previous 
results \mbox{$\S^{(0)}_{c}=\S^{(1)}_{c}=0$}, the constraints on the counterterm 
become
\begin{eqnarray}
    s^{(0)}\S^{(2)}_{c} &=& 0 \label{3.30}\\
    {\cal G}^{a}\S^{(2)}_{c} &=& 0 \label{3.31}\\
    \d_{\m}\S^{(2)}_{c} &=& 0\label{3.32}
    \end{eqnarray}
Again, the most general solution of the supersymmetry condition 
\equ{3.32}, is
\eq
\S_{c}^{(2)} = \d^{3} \S^{dim=4}_{\Phi\Pi=3}\ .
\eqn{3.33}
The situation here is a bit more involved: the most general 
$O(\theta^{2})$ candidate satisfying the ghost equation \equ{3.31}
and the supersymmetry condition 
\equ{3.32}, turns out to be
\eq
\theta^{2}\cdot\S_{c}^{(2)} =
\d^{3}\left ( \theta^{2}\cdot\S^{4}_{3} \right ) =
\d^{3} \int d^{3}x\
\left (
\theta_{\m\n}\theta^{\m\n}\S +
\theta^{\m\lambda}\theta^{\n}_{\lambda}\S_{\m\n} +
\theta^{\m\n}\theta^{\r\s}\S_{\m\n\r\s}
\right )
\eqn{3.34}
where
\begin{eqnarray}
\S &=& \int d^{3}x\ \left (
T_{1}^{[ab]cd}c^{a\r}c^{b}_{\r}c^{c\s}A^{d}_{\s} +
\a_{1}f^{abc}c^{a\r\s}c^{b}_{\r}c^{c}_{\s}
\right ) \label{3.35}\\
\S_{\m\n} &=&  \int d^{3}x\ \left (
T_{2}^{[ab]cd}c^{a}_{\m}c^{b}_{\n}c^{c\s}A^{d}_{\s} +
T_{3}^{[ab]cd}c^{a\r}c^{b}_{\r}c^{c}_{\m}A^{d}_{\n} + \right. \nonumber \\
&&\left .
+(\a_{2}d^{abc} + 
\a_{3}f^{abc})c^{a}_{\m\s}c^{b}_{\n}c^{c\s} 
\right ) \label{3.36}\\
\S_{\m\n\r\s} &=&  \int d^{3}x\ \left (
T_{4}^{(ab)cd}c^{a}_{\m}c^{b}_{\n}c^{c}_{\r}A^{d}_{\s} +
\a_{4}f^{abc}c^{a}_{\m\r}c^{b}_{\n}c^{c}_{\s}
\right )\ , \label{3.37}
\end{eqnarray}
where $c^{a}_{\m}(x)\equiv \partial_{\m}c^{a}(x)$, $\a_{i}$ are 
constants and $T_{i}^{abcd}$ are constant invariant tensors. Square 
and round brackets mean antisymmetrization and symmetrization of 
color indices, respectively.

The $s^{(0)}$ operator, which does not depend on $\theta$, does not 
mix the three sectors which form $\theta^{2}\cdot\S^{(2)}_{c}$, hence 
each of them can be studied separately
\eq
s^{(0)}\S=s^{(0)}\S_{\m\n}=s^{(0)}\S_{\m\n\r\s}=0
\eqn{3.38}
where again we used Eq. \equ{3.28}.

A careful analysis of the above BRS conditions leads to the following 
relations between the free parameters
\begin{description}
\item[sector $\theta_{\m\n}\theta^{\m\n}$]
\begin{eqnarray}
    T_{1}^{[ab]cd} &-& T_{1}^{[ab]dc} = \a_{1}f^{abp}f^{pcd} 
    \label{3.39}\\
     T_{1}^{[ab]cd} &+& T_{1}^{[ab]dc} \equiv T_{1}^{[ab](cd)}
     \ :\ \mbox{undetermined}\label{3.40}
\end{eqnarray}
\item[sector $\theta_{\m}^{\lambda}\theta_{\n\lambda}$]
\begin{eqnarray}
&& T_{2}^{[ab][cd]} + T_{3}^{[cd][ab]} - \frac{\a_{2}}{4} \left (
d^{pbd}f^{pac} - d^{pad}f^{pbc} -
d^{pbc}f^{pad} + d^{pac}f^{pbd}
\right) 
\nonumber \\
&&- \frac{\a_{3}}{2}f^{abp}f^{pcd}=0\label{3.41}
\end{eqnarray}
\eq
T_{2}^{[ab](cd)}\ ,\ T_{3}^{[ab](cd)}\ :\
\mbox{undetermined}
\eqn{3.42}
\item[sector $\theta_{\m\n}\theta_{\r\s}$]
\begin{eqnarray}
    T_{4}^{(ab)cd} &+& T_{4}^{(ab)dc} = 3\a_{4}d^{abp}d^{pcd} 
    \label{3.43}\\
     T_{4}^{(ab)cd} &-& T_{4}^{(ab)dc} \equiv T_{4}^{(ab)[cd]}
     \ :\ \mbox{undetermined}\ .\label{3.44}
\end{eqnarray}
Notice that, in order to write Eq.~\equ{3.43}, we used the relation 
\equ{2.32}, which holds only for $U(n)$ groups.
\end{description}
The three terms which form the counterterm \equ{3.34} are then
\begin{eqnarray}
\S &=& \int d^{3}x\ \left (
(\frac{\a_{1}}{2}f^{abp}f^{pcd} + T_{1}^{[ab](cd)})
c^{a\r}c^{b}_{\r}c^{c\s}A^{d}_{\s} + \right. \nonumber \\
&&\left.
+ \a_{1}f^{abc}c^{a\r\s}c^{b}_{\r}c^{c}_{\s}
\right ) \label{3.45}\\
\S_{\m\n} &=&  \int d^{3}x\ \left (
T_{2}^{[ab]cd}c^{a}_{\m}c^{b}_{\n}c^{c\s}A^{d}_{\s} +
T_{3}^{[ab]cd}c^{a\r}c^{b}_{\r}c^{c}_{\m}A^{d}_{\n} + \right. \nonumber \\
&&\left .
+(\a_{2}d^{abc} + 
\a_{3}f^{abc})c^{a}_{\m\s}c^{b}_{\n}c^{c\s} 
\right ) \label{3.46}\\
\S_{\m\n\r\s} &=&  \int d^{3}x\ \left (
(\frac{3}{2}\a_{4}d^{abp}d^{pcd} + T_{4}^{(ab)[cd]})
c^{a}_{\m}c^{b}_{\n}c^{c}_{\r}A^{d}_{\s} + \right. \nonumber \\
&&\left.
+ \a_{4}f^{abc}c^{a}_{\m\r}c^{b}_{\n}c^{c}_{\s}
\right )\ . \label{3.47}
\end{eqnarray}
The noncommutative theory, hence, starting from order $O(\theta^{2})$, not 
only breaks finiteness, but it is not even stable under radiative 
corrections, as the counterterm cannot be reabsorbed by a 
renormalization of field and parameter of the classical theory,
nor it can be expressed in terms of the Groenewald-Moyal product.

Nonetheless, we verified that, at least at $O(\theta^{2})$, 
the above nine-fold (as many are the 
free parameters) instability belongs to the nonphysical sector of the 
theory, since the unstable counterterm can be written as an exact BRS 
cocycle
\eq
\theta^{2}\cdot\S_{c}^{(2)} =
s^{(0)}\d^{3} \int d^{3}x\
\left (
\theta_{\m\n}\theta^{\m\n}\widehat\S +
\theta^{\m\lambda}\theta^{\n}_{\lambda}\widehat{\S}_{\m\n} +
\theta^{\m\n}\theta^{\r\s}\widehat{\S}_{\m\n\r\s}
\right )
\eqn{3.48}
where
\begin{eqnarray}
\widehat\S &=& \int d^{3}x\ \left (
\frac{\a_{1}}{2}f^{abc}c^{a\r}c^{b}_{\r}\partial A^{c} - \frac{1}{2} 
T_{1}^{[ab](cd)}c^{a\r}c^{b}_{\r}A^{c\s}A^{d}_{\s}
\right ) \label{3.49}\\
\widehat\S_{\m\n} &=&  - \int d^{3}x\ \left ( 
(\a_{2}d^{abc}+\a_{3}f^{abc})A^{a\r}c^{b}_{\m\r}c^{c}_{\n} 
-T_{3}^{[ab][cd]}c^{a\r}A^{b}_{\r}c^{c}_{\m}A^{d}_{\n}
\right. \label{3.50} \\
&&\left. +
(\frac{\a_{2}}{4}f^{abp}d^{pcd} +\frac{1}{2} T_{2}^{[ab](cd)})
c^{a}_{\m}c^{b}_{\n}A^{c\r}A^{d}_{\r}
+\frac{1}{2}T_{3}^{[ab](cd)}c^{a\r}c^{b}_{\r}A^{c}_{\m}A^{d}_{\n}
\right )\nonumber \\
\widehat\S_{\m\n\r\s} &=&  \int d^{3}x\ \left (
\a_{4}f^{abc}c^{a}_{\n}c^{b}_{\r}\partial_{\m}A^{c}_{\s}
-\frac{1}{2}T_{4}^{(ab)[cd]}c^{a}_{\m}c^{b}_{\n}A^{c}_{\r}A^{d}_{\s}
\right )\label{3.51}
\end{eqnarray}
Notice that, in order to write the counterterm as an exact BRS 
cocycle, we had to use the relation \equ{3.41}, which, although not attractive, 
turns out to be important to get this result.

Hence, we found that, till order $O(\theta^{2})$, the counterterm can 
be written as
\eq
\S^{(2)}_{c} = \left.
s^{(\theta)}\d^{3}\S_{\Phi\Pi=2}^{dim=0}
\right |_{O(\theta^{2})}\ ,
\eqn{3.52}
where $\S^{0}_{2}$ is a generic power series in $\theta$, which 
depends on the ghost field $c^{a}(x)$ only if differentiated. In the 
Appendix, we show that the integrated cohomology of the BRS operator 
$s^{(\theta)}$ in the ghost sectors $0$ and $+1$ is trivial, and 
hence the most general solution of  the BRS constraint \equ{3.5} is
\eq
\S^{(\theta)}_{c} = 
s^{(\theta)} \widehat\S\ .
\eqn{3.53bis}
This result, together with Eq.\equ{3.14} on the solution of the 
supersymmetry condition, leads us to conclude that, {\it to all 
orders in $\hbar$ and $\theta$}, the most general counterterm of the 
NCCS theory is indeed of the form \equ{3.52}
\eq
\S^{(\theta)}_{c} = 
s^{(\theta)}\d^{3}\S_{\Phi\Pi=2}^{dim=0}\ .
\eqn{3.54bis}
We shall comment on this in the Conclusions.

For what concerns anomaly, analogous results hold. The noncommutative 
anomaly ${\cal A}^{(\theta)}$, which has mass dimensions $+3$ and ghost 
number $+1$, must satisfy the same constraints as the counterterm
\begin{eqnarray}
    {\cal G}^{a}{\cal A}^{(\theta)} &=& 0 \label{3.53}\\
    \d_{\m}{\cal A}^{(\theta)} &=& 0 \label{3.54}\\
    s^{(\theta)}{\cal A}^{(\theta)} &=&
\left (\sum_{n=0}^{\infty}\theta^{n}\cdot s^{(n)}\right)
\left(\sum_{m=0}^{\infty}\theta^{m}\cdot {\cal A}^{(m)}\right)
=0\ ,\label{3.55}
\end{eqnarray}
and ${\cal A}^{(\theta)}$ must be closed but not exact
\eq
{\cal A}^{(\theta)}\neq s^{(\theta)} \widehat{\cal A}\ .
\eqn{3.55.1}
The first two conditions \equ{3.53} and \equ{3.54}
are solved by
\eq
{\cal A}^{(\theta)} = \d^{3}\ {\cal A}^{dim=0}_{\Phi\Pi=4}\ ,
\eqn{3.56}
and ${\cal A}^{0}_{4}$ is a local integrated functional depending on 
the ghost field only if differentiated.

The BRS condition \equ{3.55}, up to second order in $\theta$, reads
\begin{eqnarray}
    O(\theta^{0})  &:& s^{(0)}{\cal A}^{(0)} = 0 \label{3.57}\\
     O(\theta^{1})  &:& s^{(0)}{\cal A}^{(1)} +
     s^{(1)}{\cal A}^{(0)}= 0 \label{3.58}\\
 O(\theta^{2})  &:& s^{(0)}{\cal A}^{(2)} +
     s^{(1)}{\cal A}^{(1)} +  s^{(2)}{\cal A}^{(0)} = 0\ .\label{3.59} 
     \end{eqnarray}
Now, ${\cal A}^{(0)}$ and ${\cal A}^{(1)}$ are ruled out by the 
solution \equ{3.56}: the commutative theory and its noncommutative 
extension at $O(\theta)$ are not anomalous
\eq
{\cal A}^{(0)}={\cal A}^{(1)}=0\ .
\eqn{3.60}
At $O(\theta^{2})$ the most general solution of the constraints is
\begin{eqnarray}
A^{(2)}&=&\d^{3}\int d^{3}x\ \left(
T_{1}^{[ab][cd]}\theta_{\m\n}\theta^{\m\n} 
c^{a\r}c^{b}_{\r}c^{c\s}c^{d}_{\s} +
T_{2}^{[ab][cd]}\theta_{\m}^{\lambda}\theta_{\lambda\n} 
c^{a\m}c^{b\n}c^{c\r}c^{d}_{\r} +
\right. \nonumber \\
&& + \left.
T_{3}^{[ab][cd]}\theta_{\m\n}\theta_{\r\s} 
c^{a\m}c^{b\n}c^{c\r}c^{d\s}
\right )\ ,
\label{3.62}
\end{eqnarray}
which can be expresses as an exact BRS cocycle
\begin{eqnarray}
A^{(2)}&=&s^{(0)} \d^{3}\int d^{3}x\ \left(
T_{1}^{[ab][cd]}\theta_{\m\n}\theta^{\m\n} 
A^{a\r}c^{b}_{\r}c^{c\s}c^{d}_{\s} +
T_{2}^{[ab][cd]}\theta_{\m}^{\lambda}\theta_{\lambda\n} 
A^{a\m}c^{b\n}c^{c\r}c^{d}_{\r} +
\right. \nonumber \\
&& + \left.
T_{3}^{[ab][cd]}\theta_{\m\n}\theta_{\r\s} 
A^{a\m}c^{b\n}c^{c\r}c^{d\s}
\right )\ ,
\label{3.63}
\end{eqnarray}
and therefore, also at order $O(\theta)^{2}$, we explicitly checked 
that the noncommutative theory is not anomalous.

In the Appendix we show that this result holds to all orders in 
$\theta$ (and $\hbar$) as well
\eq
{\cal A}^{(\theta)}=0\ .
\eqn{anomaly}

\section{Conclusions}

In this paper we considered the noncommutative Chern-Simons theory, 
expanded in the noncommutativity parameter $\theta_{\m\n}$. This 
expansion covers the whole theory, since we showed that a 
non-analytical expansion is not allowed in this case.
Therefore, we faced with a double expansion: a quantum expansion in 
$\hbar$ and a noncommutative one in $\theta_{\mu\nu}$.

We gave the most general expression for the counterterm in 
\equ{3.54bis}. Due to the presence of a generic functional 
$\S^{0}_{2}$, the counterterm depends on an infinite number of free 
parameters, and, consequently, represents an infinite set of unstable 
radiative corrections to the classical noncommutative action. 
Moreover, the counterterm cannot be written in terms of the 
Groenewald-Moyal star product, which therefore turns out to be 
unstable under radiative corrections. This seems to indicate that the 
quantum theory loses its link to an underlying noncommutative 
structure of spacetime.

The optimistic counterpart, is that all the above considerations are 
confined to the nonphysical sector of the quantum theory, which is 
also anomaly free. The bulk of the theory maintains unaltered the good 
properties of the commutative one: the $\b$ function of the 
noncommutative Chern-Simons coupling constant vanishes. We proved also that the noncommutative parameter $\theta$ is 
a nonphysical coupling 
constant, since, like what happens for gauge parameters, 
the fact that the counterterm is an exact BRS cocycle implies 
that 
\eq
\theta_{\m\n}\frac{\partial\G^{(\theta)}}{\partial\theta_{\m\n}}
= s^{(\theta)}
\int d^{3}x\ \D\cdot\G^{(\theta)}\ ,
\eqn{theta}
where $\G^{(\theta)}$ is the quantum noncommutative action, and 
$\D\cdot\G^{(\theta)}$ is a quantum insertion.

These results suggest to infer that the noncommutative extension 
leads to a quantum field theory which is consistent as long as the 
physical sector is concerned, while it is less meaningful for the 
part of the theory which determines the anomalous dimensions.

We stress that the job has been greatly simplified in the case we 
considered: a three dimensional topological field theory whose 
dependence on the noncommutative parameter is completely analytic, 
and whose set of symmetries, in particular the vector 
supersymmetry, allowed us to study thoroughly higher orders in both 
$\hbar$ and $\theta$.

It would be extremely interesting to make analogous investigations in 
more physically relevant quantum field theories, like for instance 
Yang-Mills, or also Maxwell theory.

%
%
%
%
\appendix

\section{All Orders Results}

\subsection{Anomaly}

We want to show that the general solution of the Wess-Zumino 
consistency condition on the 
anomaly
\eq
s^{(\theta)}{\cal A}^{(\theta)}= 0
\eqn{a1}
is an exact BRS cocycle
\eq
{\cal A}^{(\theta)} = s^{(\theta)}\widehat{\cal A}^{(\theta)}\ .
\eqn{a2}
Where ${\cal A}^{\theta}$ and $\widehat{\cal A}^{\theta}$ are 
integrated local functionals with mass dimension $+3$ and ghost 
number $+1$, which can be expressed, like the BRS 
operator $s^{(\theta)}$, as power series in $\theta$
\begin{eqnarray}
    {\cal A}^{(\theta)} &=& \sum_{n=0}^{\infty}\theta^{n}\cdot 
    {\cal A}^{(n)} \label{a3} \\
    s^{(\theta)} &=& \sum_{n=0}^{\infty}\theta^{n}\cdot 
    s^{(n)} \label{a4}\ ,
    \end{eqnarray}
where the same prescription as in \equ{3.1} is adopted concerning the ``dot'' 
products.    

The global equation \equ{a1}, written on local forms, reads
\eq
s^{(\theta)}{\cal A}^{3}_{1}(x) + d {\cal A}^{2}_{2}(x)=0\ ,
\eqn{a5}
where ${\cal A}^{p}_{q}(x)$ is a local p-form with ghost number $q$, 
and $d$ is the exterior derivative.

In Section 3 we showed that, up to second order in $\theta$, it holds
\eq
({\cal A}^{3}_{1})^{(2)} = 
\left (s^{(\theta)}{\cal A}^{3}_{0} +d {\cal A}^{2}_{1}\right )^{(2)} =
s^{(0)} ({\cal A}^{3}_{0})^{(2)} +
s^{(1)} ({\cal A}^{3}_{0})^{(1)} +
s^{(2)} ({\cal A}^{3}_{0})^{(0)} +d ({\cal A}^{2}_{1})^{(2)} \ ,
\eqn{a6}
and we neglect that, in particular, $({\cal A}^{3}_{0})^{(0)}=({\cal 
A}^{3}_{0})^{(1)}=0$, due to the ghost equation.

The proof develops by induction: we assume that, up to order $n-1$, it 
holds
\begin{eqnarray}
({\cal A}^{3}_{1})^{(n-1)} &=&
\left (s^{(\theta)}{\cal A}^{3}_{0} + d {\cal A}^{2}_{1}\right 
)^{(n-1)} 
\label{a7} \\
&=& s^{(0)} ({\cal A}^{3}_{0})^{(n-1)} +
s^{(1)} ({\cal A}^{3}_{0})^{(n-2)} + .. +
s^{(n-2)} ({\cal A}^{3}_{0})^{(1)} \nonumber \\ 
&& +
s^{(n-1)} ({\cal A}^{3}_{0})^{(0)} +
d ({\cal A}^{2}_{1})^{(n-1)} \nonumber\ ,
\end{eqnarray}
and we want to show that, also at the next order, 
\begin{eqnarray}
({\cal A}^{3}_{1})^{(n)} &=&
\left (s^{(\theta)}{\cal A}^{3}_{0} + d {\cal A}^{2}_{1}\right )^{(n)} 
\label{a8} \\
&=& s^{(0)} ({\cal A}^{3}_{0})^{(n)} +
s^{(1)} ({\cal A}^{3}_{0})^{(n-1)} + .. +
s^{(n-1)} ({\cal A}^{3}_{0})^{(1)} \nonumber \\
&& +
s^{(n)} ({\cal A}^{3}_{0})^{(0)} +
d ({\cal A}^{2}_{1})^{(n)} \nonumber\ ,
\end{eqnarray}
where $({\cal A}^{3}_{1})^{(n)}$ satisfies the equation \equ{a5} at the 
order $O(\theta^{n})$, that is
\eq
s^{(0)} ({\cal A}^{3}_{1})^{(n)} +
s^{(1)} ({\cal A}^{3}_{1})^{(n-1)} + .. +
s^{(n-1)} ({\cal A}^{3}_{1})^{(1)} +
s^{(n)} ({\cal A}^{3}_{1})^{(0)} +
d ({\cal A}^{2}_{2})^{(n)} =0\ .
\eqn{a9}
Substituting \equ{a7} in \equ{a9}, and grouping the same 
$\theta$-powers of ${\cal A}^{3}_{0}$, we get
\begin{eqnarray}
&&
s^{(0)}({\cal A}^{3}_{1})^{(n)} + \nonumber \\
&&
s^{(1)}s^{(0)}({\cal A}^{3}_{0})^{(n-1)} + \nonumber \\
&&
((s^{(1)})^{2} + s^{(2)}s^{(0)})({\cal A}^{3}_{0})^{(n-2)} + .. + \label{a10}\\
&&
(s^{(1)}s^{(n-2)} + s^{(2)}s^{(n-3)} + .. + 
s^{(n-2)}s^{(1)}+s^{(n-1)}s^{(0)})
({\cal A}^{3}_{0})^{(1)} +  \nonumber \\
&&
(s^{(1)}s^{(n-1)} + s^{(2)}s^{(n-2)} + .. + 
s^{(n-1)}s^{(1)}+s^{(n)}s^{(0)})
({\cal A}^{3}_{0})^{(0)} + \nonumber \\
&&
d (\widehat{\cal A}^{2}_{2})^{(n)}
=0\ .\nonumber
\end{eqnarray}
Using the nilpotency relation at $O(\theta^{n})$
\eq
s^{(0)}s^{(n)} + s^{(1)}s^{(n-1)} + .. +
s^{(n-1)}s^{(1)} + s^{(n)}s^{(0)} =0\ ,
\eqn{a11}
the Eq.\equ{a10} writes
\begin{eqnarray}
&&s^{(0)} \left (
({\cal A}^{3}_{1})^{(n)} -
s^{(1)}({\cal A}^{3}_{0})^{(n-1)} - 
s^{(2)}({\cal A}^{3}_{0})^{(n-2)} - .. -
s^{(n-1)}({\cal A}^{3}_{0})^{(1)} - \right . \nonumber \\
&& \left. s^{(n)}({\cal A}^{3}_{0})^{(0)} 
\right ) + d (\widehat{\cal A}^{2}_{2})^{(n)} = 0\label{a12}\ ,
\end{eqnarray}
which is an equation of cohomology modulo-$d$ for the ordinary, 
commutative, nilpotent, BRS operator $s^{(0)}$. 

At this point, we proceed as usual, applying $s^{(0)}$ to both sides 
of \equ{a12}, using the anticommutation relation  $\{s^{(0)},d\}=0$, 
the nilpotency $d^{2}=0$, and the fact that the cohomology of $d$ 
is empty, and we get the following descent equations 
\cite{Piguet:1995er}, which hold for each $O(\theta^{n})$
\begin{eqnarray}
s^{(0)}(\widehat{\cal A}^{2}_{2})^{(n)} + d ({\cal A}^{1}_{3})^{(n)} &=& 0
\label{a13}\\
s^{(0)}({\cal A}^{1}_{3})^{(n)} + d ({\cal A}^{0}_{4})^{(n)} &=& 0
\label{a14}\\
s^{(0)}({\cal A}^{0}_{4})^{(n)} &=& 0
\label{a15}\ .
\end{eqnarray}
The last equation \equ{a15} is a local cohomology equation for 
$s^{(0)}$, and we know that it is formed by odd polynomials in the 
undifferentiated ghost $c^{a}(x)$ \cite{Piguet:1995er}
\eq
{\cal H}(s^{(0)}) = {\cal P}_{odd}(c)\ .
\eqn{a16}
Therefore, in the sector with even ghost number, the local cohomology 
of $s^{(0)}$ is empty, and the solution of \equ{a15} is
\eq
({\cal A}^{0}_{4})^{(n)} = s^{(0)} ({\cal A}^{0}_{3})^{(n)}\ .
\eqn{a17}
Now, it is easy to mount the descent equations, remembering \equ{a16} 
and the fact that the ghost field has vanishing mass dimensions. We 
get
\eq
(\widehat{\cal A}^{2}_{2})^{(n)} = s^{(0)}({\cal A}^{2}_{1})^{(n)} +
d ({\cal A}^{1}_{2})^{(n)}\ ,
\eqn{a18}
which, substituted in \equ{a12}, transforms the problem of cohomology 
modulo-$d$ into a problem of local cohomology
\eq
s^{(0)}\left (
({\cal A}^{3}_{1})^{(n)} - 
s^{(1)}({\cal A}^{3}_{0})^{(n-1)} - .. -
s^{(n-1)}({\cal A}^{3}_{0})^{(1)}-
s^{(n)}({\cal A}^{3}_{0})^{(0)} - d({\cal A}^{2}_{1})^{(n)}
\right) = 0\ ,
\eqn{a19}
which is solved by \equ{a8}, which was our aim. 

Therefore, at any order in $\theta$, the solution of the Wess-Zumino 
consistency condition \equ{a1} is only the trivial one \equ{a2}, 
and the noncommutative Chern-Simons theory is not anomalous.

\subsection{Counterterm}

The computation of the counterterm follows the same steps as the 
anomaly. We want to solve the BRS constraint on the counterterm
\eq
s^{(\theta)}\S^{(\theta)}= 0\ ,
\eqn{a20}
where $\S^{(\theta)}$ is a local integrated functional with mass 
dimensions $+3$ and ghost number $0$, which can be expanded in power 
series of $\theta$
\eq
\S^{(\theta)} = \sum_{n=0}^{\infty}\theta^{n}\cdot 
    \S^{(n)}\ .
\eqn{a21}
Written in terms of differential forms, \equ{a20} reads
\eq
s^{(\theta)}\S^{3}_{0}(x)+d\S^{2}_{1}(x)=0\ .
\eqn{a22}
In Section 3 we showed that, up to second order in $\theta$,
\eq
(\S^{3}_{0})^{(2)} =
(s^{\theta}\S^{3}_{-1} + d \S^{2}_{0})^{(2)}\ .
\eqn{a23}
By induction, we suppose that, up to $O(\theta^{n-1)}$, 
\eq
(\S^{3}_{0})^{(n-1)} =
\left ( s^{(\theta)}\S^{3}_{-1} + d \S^{2}_{0}\right )^{(n-1)}\ ,
\eqn{a24}
and we want to show that, at the next order, 
\eq
(\S^{3}_{0})^{(n)} =
\left ( s^{(\theta)}\S^{3}_{-1} + d \S^{2}_{0}\right )^{(n)}\ ,
\eqn{a25}
where $(\S^{3}_{0})^{(n)}$ is constrained to satisfy \equ{a22}. 
Substituting \equ{a24} 
in \equ{a22}, reordering terms and using the 
nilpotency relation \equ{a11}, in an analogous way to the anomaly case, 
we land on a local cohomology modulo-$d$ problem for the ordinary 
BRS operator $s^{(0)}$:
\eq
s^{(0)}\left (
(\S^{3}_{0})^{(n)} - s^{(1)}(\S^{3}_{-1})^{(n-1)}-..-
s^{(n)}(\S^{3}_{-1})^{(0)}\right)+d(\widehat\S^{2}_{1})^{(n)}=0\ ,
\eqn{a26}
which yields the descent equations
\begin{eqnarray}
s^{(0)}(\widehat\S^{2}_{1})^{(n)} + d (\S^{1}_{2})^{(n)} &=& 0 
\label{a27} \\
s^{(0)}(\S^{1}_{2})^{(n)} + d (\S^{0}_{3})^{(n)} &=& 0 
\label{a28} \\
s^{(0)}(\S^{0}_{3})^{(n)} &=& 0\ . \label{a29}
\end{eqnarray}
The last \equ{a29} is a local cohomology equation for the ordinary 
BRS operator, for which the result \equ{a16} holds. This time the 
ghost charge of the $0$-form $(\S^{0}_{3})^{(n)}$ is odd, but, since 
its mass dimension is $2n$, the local cohomology also in this case is 
vanishing, since no dimensionful quantities can be constructed with 
polynomials of undifferentiated ghost fields. Hence
\eq
(\S^{0}_{3})^{(n)} = s^{(0)}(\S^{0}_{2})^{(n)}\ .
\eqn{a30}
The descent equations are easily mounted, up to
\eq
(\widehat\S^{2}_{1})^{(n)}=s^{(0)}(\S^{2}_{0})^{(n)}+d(\S^{1}_{1})^{(n)}\ ,
\eqn{a31}
which, substituted in \equ{a26}, leads to a local cohomology problem, 
solved by \equ{a25}. 

Therefore, we have shown that the integrated cohomology of 
$s^{(\theta)}$ is empty in all sectors $O(\theta^{n})$, with $n\geq 1$.


\begin{thebibliography}{999}
\bibitem{Douglas:2001ba}
  M.~R.~Douglas and N.~A.~Nekrasov,
  {\it Noncommutative field theory},
  Rev.\ Mod.\ Phys.\  {\bf 73}, 977 (2001)
  [arXiv:hep-th/0106048].
\bibitem{Szabo:2001kg}
  R.~J.~Szabo,
  {\it Quantum field theory on noncommutative spaces},
  Phys.\ Rept.\  {\bf 378}, 207 (2003)
  [arXiv:hep-th/0109162].
\bibitem{Blasi:2005bk}
  A.~Blasi, N.~Maggiore and M.~Montobbio,
  {\it Instabilities of noncommutative two dimensional BF model},
  Mod.\ Phys.\ Lett.\ A {\bf 20}, 2119 (2005)
  [arXiv:hep-th/0504218].
\bibitem{Blasi:2005vf}
  A.~Blasi, N.~Maggiore and M.~Montobbio,
  {\it Noncommutative two dimensional BF model},
  Nucl.\ Phys.\ B {\bf 740}, 281 (2006)
  [arXiv:hep-th/0512006].
\bibitem{Gomis:1995jp}
  J.~Gomis and S.~Weinberg,
  {\it Are Nonrenormalizable Gauge Theories Renormalizable?},
  Nucl.\ Phys.\  B {\bf 469}, 473 (1996)
  [arXiv:hep-th/9510087].
\bibitem{Blasi:1998ph}
  A.~Blasi, N.~Maggiore, S.~P.~Sorella and L.~C.~Q.~Vilar,
  {\it Renormalizability of nonrenormalizable field theories},
  Phys.\ Rev.\  D {\bf 59}, 121701 (1999)
  [arXiv:hep-th/9812040].
\bibitem{Blasi:2006gq}
  A.~Blasi and N.~Maggiore,
  {\it General solution of vector supersymmetry},
  Class.\ Quant.\ Grav.\  {\bf 24}, 645 (2007)
  [arXiv:hep-th/0606124].
\bibitem{Delduc:1989ft}
  F.~Delduc, F.~Gieres and S.~P.~Sorella,
  {\it Supersymmetry Of The D = 3 Chern-Simons Action In The Landau Gauge},
  Phys.\ Lett.\ B {\bf 225}, 367 (1989).
\bibitem{Maggiore:1991aa}
  N.~Maggiore and S.~P.~Sorella,
  {\it Finiteness of the topological models in the Landau gauge},
  Nucl.\ Phys.\ B {\bf 377}, 236 (1992).
\bibitem{Birmingham:1991rh}
  D.~Birmingham and M.~Rakowski,
  {\it Vector supersymmetry in topological field theory},
  Phys.\ Lett.\ B {\bf 269}, 103 (1991).
\bibitem{Blasi:1990xz}
  A.~Blasi, O.~Piguet and S.~P.~Sorella,
  {\it Landau Gauge And Finiteness},
  Nucl.\ Phys.\ B {\bf 356}, 154 (1991).
\bibitem{MacFarlane:1968vc}
  A.~J.~MacFarlane, A.~Sudbery and P.~H.~Weisz,
  {\it On Gell-Mann's Gamma Matrices, D Tensors And F Tensors, Octets, And
  Parametrizations Of SU(3)},
  Commun.\ Math.\ Phys.\  {\bf 11}, 77 (1968).
\bibitem{Piguet:1995er}
  O.~Piguet and S.~P.~Sorella,
  {\it Algebraic renormalization: Perturbative renormalization, symmetries and
  anomalies},
  Lect.\ Notes Phys.\  {\bf M28}, 1 (1995).
\bibitem{Blasi:1989mw}
  A.~Blasi and R.~Collina,
  {\it Finiteness Of The Chern-Simons Model In Perturbation Theory},
  Nucl.\ Phys.\ B {\bf 345}, 472 (1990).
\bibitem{Delduc:1990je}
  F.~Delduc, C.~Lucchesi, O.~Piguet and S.~P.~Sorella,
  {\it Exact Scale Invariance Of The Chern-Simons Theory In The Landau Gauge},
  Nucl.\ Phys.\ B {\bf 346}, 313 (1990).
\bibitem{Dixon:1991wi}
  J.~A.~Dixon,
  {\it Calculation of BRS cohomology with spectral sequences},
  Commun.\ Math.\ Phys.\  {\bf 139}, 495 (1991).
\end{thebibliography}
\end{document}